\documentclass[oneside,twocolumn,aps,prl,preprintnumbers,bibnotes10pt,superscriptaddress,amsmath,amssymb]{revtex4-2}
\usepackage{graphicx,subfigure,wrapfig}
\usepackage{float}
\usepackage{color}
\usepackage[dvipsnames]{xcolor}
\usepackage{epstopdf}
\usepackage{amsmath}
\usepackage{braket}
\usepackage{amsthm}
\usepackage{amssymb}
\usepackage{amsfonts}
\usepackage{graphicx,subfigure}
\usepackage{txfonts}
\usepackage{ulem}
\usepackage{mathtools}
\usepackage{bm}
\usepackage{bbm}
\usepackage{hyperref}
\usepackage{natbib}
\usepackage{comment}
\usepackage{cancel}

\newcommand{\remove}[1]{}                                                 

\definecolor{giu}{rgb}{1,0.7,0.7}

\newcommand{\be}{\begin{equation}}
\newcommand{\ee}{\end{equation}}
\newcommand{\beq}{\begin{eqnarray}}
\newcommand{\eeq}{\end{eqnarray}}
\newcommand{\vect}[1]{\bm{#1}}

\begin{document}

\title{Self-induced Josephson oscillations and self-trapping \\ in a supersolid dipolar quantum gas}

\author{Beatrice Donelli}
\affiliation{Istituto Nazionale di Ottica, Consiglio Nazionale delle Ricerche (INO-CNR), Largo Enrico Fermi 2, 50125 Firenze, Italy}
\affiliation{European Laboratory for Nonlinear Spectroscopy (LENS), Via Nello Carrara 1, 50019 Sesto Fiorentino, Italy}

\author{Nicolò Antolini}
\affiliation{European Laboratory for Nonlinear Spectroscopy (LENS), Via Nello Carrara 1, 50019 Sesto Fiorentino, Italy}
\affiliation{Istituto Nazionale di Ottica, Consiglio Nazionale delle Ricerche (INO-CNR), Via Moruzzi 1, 56124 Pisa, Italy}

\author{Giulio Biagioni}
\affiliation{European Laboratory for Nonlinear Spectroscopy (LENS), Via Nello Carrara 1, 50019 Sesto Fiorentino, Italy}
\affiliation{Université Paris-Saclay, Institut d’Optique Graduate School, CNRS, Laboratoire Charles Fabry, 91127, Palaiseau, France}

\author{Marco Fattori}
\affiliation{Istituto Nazionale di Ottica, Consiglio Nazionale delle Ricerche (INO-CNR), Largo Enrico Fermi 2, 50125 Firenze, Italy}
\affiliation{European Laboratory for Nonlinear Spectroscopy (LENS), Via Nello Carrara 1, 50019 Sesto Fiorentino, Italy}
\affiliation{Dipartimento di Fisica e Astronomia, Università degli Studi di Firenze, Via G. Sansone 1, 50019 Sesto Fiorentino, Italy}

\author{Andrea Fioretti}
\affiliation{Istituto Nazionale di Ottica, Consiglio Nazionale delle Ricerche (INO-CNR), Via Moruzzi 1, 56124 Pisa, Italy}

\author{Carlo Gabbanini}
\affiliation{Istituto Nazionale di Ottica, Consiglio Nazionale delle Ricerche (INO-CNR), Via Moruzzi 1, 56124 Pisa, Italy}

\author{Massimo Inguscio}
\affiliation{European Laboratory for Nonlinear Spectroscopy (LENS), Via Nello Carrara 1, 50019 Sesto Fiorentino, Italy}
\affiliation{Dipartimento di Ingegneria, Università Campus Bio-Medico di Roma, Via Alvaro Del Portillo 21, 00128 Roma, Italy}

\author{Luca Tanzi}
\affiliation{Istituto Nazionale di Ottica, Consiglio Nazionale delle Ricerche (INO-CNR), Largo Enrico Fermi 2, 50125 Firenze, Italy}
\affiliation{European Laboratory for Nonlinear Spectroscopy (LENS), Via Nello Carrara 1, 50019 Sesto Fiorentino, Italy}

\author{Giovanni Modugno}
\affiliation{European Laboratory for Nonlinear Spectroscopy (LENS), Via Nello Carrara 1, 50019 Sesto Fiorentino, Italy}
\affiliation{Istituto Nazionale di Ottica, Consiglio Nazionale delle Ricerche (INO-CNR), Via Moruzzi 1, 56124 Pisa, Italy}
\affiliation{Dipartimento di Fisica e Astronomia, Università degli Studi di Firenze, Via G. Sansone 1, 50019 Sesto Fiorentino, Italy}

\author{Augusto Smerzi}
\email{augusto.smerzi@ino.it}
\affiliation{Istituto Nazionale di Ottica, Consiglio Nazionale delle Ricerche (INO-CNR), Largo Enrico Fermi 2, 50125 Firenze, Italy}
\affiliation{European Laboratory for Nonlinear Spectroscopy (LENS), Via Nello Carrara 1, 50019 Sesto Fiorentino, Italy} 
\affiliation{QSTAR, Largo Enrico Fermi 2, 50125 Firenze, Italy.}

\author{Luca Pezz$\grave{\text{e}}$}
\email{luca.pezze@ino.it}
\affiliation{Istituto Nazionale di Ottica, Consiglio Nazionale delle Ricerche (INO-CNR), Largo Enrico Fermi 2, 50125 Firenze, Italy}
\affiliation{European Laboratory for Nonlinear Spectroscopy (LENS), Via Nello Carrara 1, 50019 Sesto Fiorentino, Italy} 
\affiliation{QSTAR, Largo Enrico Fermi 2, 50125 Firenze, Italy.}

\begin{abstract}
The Josephson effect characterizes superfluids and superconductors separated by a weak link, the so-called Josephson junction. 
A recent experiment has shown that Josephson oscillations can be observed also in a supersolid, where the weak link is not due to an external barrier, but is self-induced by interparticle interactions. 
Here we show theoretically that supersolids -- despite their self-induced character -- feature all the standard properties of bosonic Josephson junction arrays, including macroscopic quantum self-trapping. 
We focus on the harmonically trapped  dipolar supersolids of interest for current experiments, and show that they can be described with a generalized Josephson model that takes into account spatial inhomogeneities.
Our work shades new light on the dynamics of supersolids and opens the way to the study of a novel class of Josephson junctions.
\end{abstract}

\maketitle

\textit{Introduction ---}
The Josephson effect~\cite{JosephsonPL1962} is a paradigmatic manifestation of macroscopic quantum coherence.
It has been observed in weakly-linked superconductors~\cite{BaroneBOOK} and  superfluids~\cite{ReviewPackardSF}, and in ultracold quantum gases with external potential barriers~\cite{CataliottiSCIENCE2001, AlbiezPRL2005, ValtolinaSCIENCE2015, SpagnolliPRL2017, KwonSCIENCE2020}.
Nonlinearities associated to interparticle interactions in weakly coupled Bose-Einstein condensates (BECs) enrich the Josephson physics providing different regimes of oscillations~\cite{SmerziPRL1997,RaghavanPRA1999}: 
small amplitude oscillations (SAO) of density and phase around the equilibrium, and macroscopic quantum self trapping (MQST) where the phase increases linearly while current oscillates around a non-zero value.
Supersolids are a long-sought phase of matter~\cite{GrossPRL1957,AndreevJETP1969,ChesterPRA1970,leggett_1970}, only recently discovered in dipolar quantum gases~\cite{TanziPRL2019,SS_Pfau,SS_Ferlaino}, which feature a spontaneous spatial modulation of the macroscopic wavefunction due only to internal forces.
The presence of a spontaneous density modulations suggests that the
Josephson effect should be intrinsically present in supersolids.
%
The supersolid flow through an obstacle has been studied in Ref.~\cite{KunimiPRB2011}, while the self-induced Josephson dynamics of a dipolar Bose gas in a torus has been investigated in Ref.~\cite{AbadEPL2011}.
The modelization of supersolids as an array of Josephson junctions has been addressed to study the rephasing dynamics after a quench~\cite{Ilzhfer2021}, quantum fluctuations~\cite{BuhlerPRR2023} and excitation spectroscopy~\cite{PlattARXIV}.
Recently, an experiment has demonstrated that Josephson SAO can indeed be excited in a small dipolar supersolid, and that their modeling can give access to key properties such as the superfluid fraction~\cite{biagioni2023subunity}.
However, how general and robust is the phenomenon? How well it can be mapped to the standard Josephson effect? 
The self induced nature of the weak links makes these questions highly non-trivial.
For instance, the low-energy oscillation mode in the harmonic trap, the so-called Goldstone mode, may affect the stability of Josephson oscillations and prevent the onset of MQST.

\begin{figure}[b!]
    \centering \includegraphics[width=\linewidth]{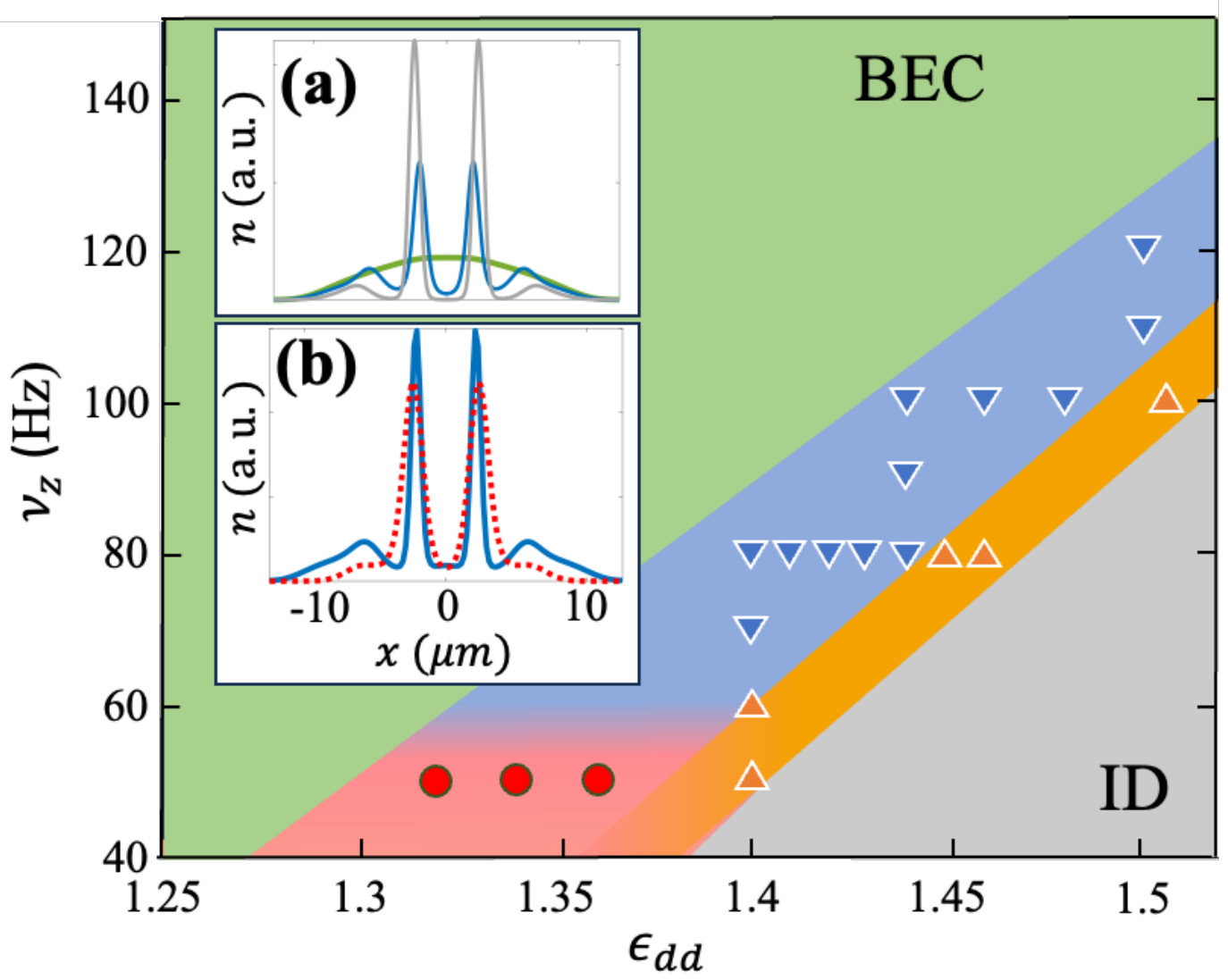}
    \caption{Dynamical phase diagram as a function of $\epsilon_{dd}$ and $\nu_z$, highlighting distinct  regimes.  
    Green and grey regions mark the BEC and ID phases, respectively~\cite{supp}. 
    Supersolid regions are denoted by colored regions corresponding to different dynamical behaviors.
    Symbols refer to numerical simulations: blue downward triangles (SAO), orange upward triangles (MQST), and red circles (MO). 
    The insets depict the ground-state density along $x$ (integrated over $y$ and $z$).  
    Inset (a) illustrates the effect of increasing $\epsilon_{dd}$ on the density contrast $n$, with examples for BEC ($\epsilon_{dd}=1.36$, green line), supersolid ($\epsilon_{dd}=1.40$, blue), and ID ($\epsilon_{dd}=1.47$, gray) at $\nu_z = 80$ Hz.
    Inset (b) shows the change in relative cluster populations with varying $\nu_z$, for $\epsilon_{dd}=1.50$, $\nu_z=120$ Hz (solid blue line) and $\epsilon_{dd}=1.32$, $\nu_z=50$ Hz (dotted red line).}
    \label{fig:phasediagram}
\end{figure}

\begin{figure*}[ht!]
    \centering
    \includegraphics[width=1\linewidth]{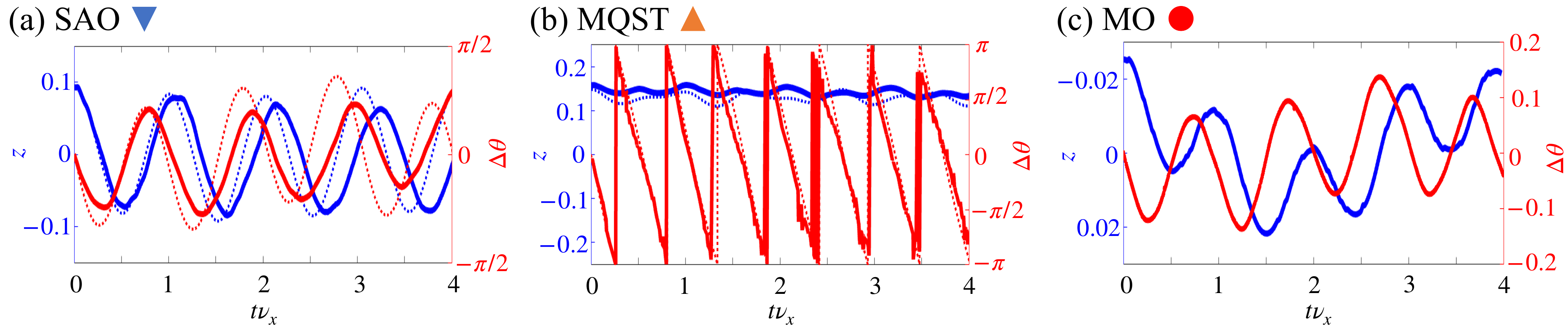}
    \caption{Examples of time evolution of the population imbalance $z(t)$ (blue lines) and the phase difference $\Delta\theta(t)$ (red lines) for the different regimes shown in the phase diagram of Fig.~\ref{fig:phasediagram}:
    SAO (a), for $\epsilon_{dd}=1.41$, $\nu_z=80$ Hz, and $z_0=0.1$; 
    MQST (b), for $\epsilon_{dd}=1.46$, $\nu_z=80$ Hz, and $z_0=0.16$;
    MO (c), for $\epsilon_{dd}=1.32$, $\nu_z=50$ Hz, and $z_0=0.03$.
    Thick lines are numerical results from the eGPE.
    Dotted lines in panels (a) and (b) are obtained from the theoretical model.}
    \label{fig:Josephsondynamics}
\end{figure*}

In this manuscript, we demonstrate theoretically that both Josephson SAO and MQST can be sustained by the supersolids in a broad regime of parameters of the dipolar supersolid, beyond the specific settings of the recent experiments~\cite{biagioni2023subunity}.
We perform numerical simulations based on the extended Gross-Pitaevskii equation (eGPE).
The numerical results are captured by an analytical
model that takes into account multiple density clusters and the spatial inhomogeneity of the harmonically-trapped supersolid. 
This allows us to clearly identify the Josephson oscillation frequency, due to effective tunneling and interaction, the SAO-to-MQST sharp transition, as well as the onset of multimode oscillations (MOs).
The latter set in for very small supersolids due to a strong breaking of the parity symmetry in the harmonic potential that results in a coupling to the Goldstone mode. 
The rich phase diagram of Fig.~\ref{fig:phasediagram} provides an overview of our findings.
Our study extends the Josephson effect beyond the traditional superfluids and superconductors to supersolids, which represent a new class of Josephson-junction arrays showing novel phenomena due to the interplay of the interparticle interactions and the self-induced nature of the weak links.

\textit{Model and Results ---} 
Our study is based on the numerical solutions of the 3D eGPE in an harmonic trap of frequencies $\nu_{x,y,z}$~\cite{supp}.
It includes contact and dipolar interaction, as well as Lee-Huang-Yang corrections~\cite{Fischer2006,LimaPelster2011,LimaPelster2012}.
This approach~\cite{Petrov2015_eGPE,WchtlerPRA2016, Roccuzzo2019,RoccuzzoPRL2020,TengstrandPRA2021} has demonstrated excellent agreement between theory and experiments~\cite{TanziNature2019, Tanzi2021, SS_Ferlaino, SS_Pfau, Ilzhfer2021,CabreraScience2018, SemeghiniPRL2018,biagioni2023subunity}.
We consider experimentally-relevant parameters~\cite{TanziPRL2019}:  $N_a=35000$ $^{162}$Dy atoms, $\nu_x = 20 \, \text{Hz}$ and $\nu_y= 50 \, \text{Hz}$, while tuning the longitudinal trap frequency $\nu_z$ and the ratio $\epsilon_{dd}=a_{dd}/a_s$ between the contact and the dipolar scattering lengths, $a_{s}$ and $a_{dd}$, respectively.
Changing $\epsilon_{dd}$ modifies the contrast of density modulations~\cite{TanziPRL2019,SS_Ferlaino,SS_Pfau,BiagioniPRX2022,Alaa2022}, see inset (a) of Fig.~\ref{fig:phasediagram}, allowing to address a standard Bose-Einstein condensate regime (BEC, green region), an intermediate supersolid phase made of four clusters (labeled as $j=1,...,4$ from left to right), and an incoherent droplet regime (ID, gray region)~\cite{supp}.
When decreasing $\nu_z$, inset (b), the supersolid lattice spacing increases, depleting the population of the lateral clusters in the ground state.
Following Ref.~\cite{biagioni2023subunity}, we first determine the ground state of the eGPE by superimposing to the harmonic trap along the $x$ axis a sinusoidal optical lattice with a half-period matching the distance between the supersolid clusters~\cite{nota1}. 
At $t=0$, we switch off the lattice potential and study 
the dynamical evolution of the system by tracking the population $N_j(t)$ and phase $\theta_j(t)$ of each cluster~\cite{note1}.
We focus on the relative population, $z(t) = \tfrac{N_{3}(t)-N_{2}(t)}{N_{3}(t)+N_{2}(t)}$, and phase, $\Delta\theta(t) = \theta_{3}(t)-\theta_{2}(t)$, between the two central clusters~\cite{note1}.
The initial excitation causes a shift of the the center of mass of the system.
In a standard BEC, this would trigger a dipole mode of frequency $\nu_x$. 
In contrast, in the supersolid, our analysis reveals a rich phase diagram with three distinct dynamical regimes: 
i) \textit{SAO} [blue region in Fig.~\ref{fig:phasediagram}.] 
A small population imbalance causes $z$ and $\Delta\theta$ to oscillate symmetrically around zero with the same frequency, exhibiting the characteristic Josephson $\pi/2$ phase shift. 
A characteristic example of SAO is shown in Fig.~\ref{fig:Josephsondynamics}(a).
ii) \textit{MQST} [orange region in Fig.~\ref{fig:phasediagram}.]. 
For fixed, relatively small $z_0 \equiv z(t=0)$, and sufficiently large $\epsilon_{dd}$, we find a regime characterized by oscillation of $z$ around a nonzero value and a running phase difference spanning the entire interval $[0,2\pi]$, see Fig.~\ref{fig:Josephsondynamics}(b).
This regime can be also accessed, for a fixed $\epsilon_{dd}$, by increasing the initial population imbalance above a critical value $z_c$, that we identify below. 
The MQST has not been experimentally observed yet in a supersolid, likely because in the experiment~\cite{biagioni2023subunity} the dynamics is triggered with a phase imprinting protocol.
iii) MOs [red region in Fig.~\ref{fig:phasediagram})]. 
In this regime, accessed at sufficiently low $\nu_z$ the coupling between the Josephson and the low-energy Goldstone mode of the supersolid~\cite{PfauNatureGoldstone} is enhanced, resulting in current-phase oscillations characterized by multiple frequencies.  
To contrast the shift of the center of mass, the Goldstone mode populates dynamically the lateral clusters resulting in an oscillation with a frequency lower than $\nu_x$, which superposes with the Josephson oscillations. 
In Fig.~\ref{fig:Josephsondynamics}(c) we clearly observe, at $\nu_z = 50$ Hz, population-phase oscillation with two frequencies, even for a very small population imbalance.   

For an analytical description of the system's dynamics, we model the supersolid as a linear array of four weakly-connected clusters, each identified by $\psi_j(t) = \sqrt{N_j(t)} e^{-i\theta_j(t)}$~\cite{biagioni2023subunity}.
We write a set of four coupled equations ($j=1,...,4$)
\begin{equation}\label{eq:modeljmode}
    i\frac{\partial\psi_j}{\partial t} =(E_j+U_jN_j)\psi_j-K_{j,j-1}\psi_{j-1}-K_{j,j+1}\psi_{j+1},
\end{equation}
where $U_j$ is the on-site interaction and $K_{j,j-1}$ is the coupling coefficient between clusters $j$ and $j-1$ ($K_{1,0}=K_{5,4}=0$ at the borders). 
The terms $E_j$ account for the energy offset of the $j$th cluster due to the external trapping.
In particular, we consider a symmetric clusters array with $E_1=E_4$, $E_2=E_3$, $U_1=U_4$ and $U_2=U_3$.
Under the condition 
\begin{equation}\label{eq:HpCurrents}
    \Dot{N}_{3}-\Dot{N}_{2} = \alpha \left(\Dot{N_{1}}-\Dot{N_{4}}\right),
\end{equation}
where $\alpha$ is a constant, Eqs.~\eqref{eq:modeljmode} give \cite{supp}
\begin{subequations}\label{eq:jj}
\begin{align}
    \Dot{z} &= 2K \frac{\alpha}{\alpha-1}\sqrt{1-z^2}\sin(\Delta\theta) \label{jj_z} \\
    \Dot{\Delta\theta} &= -U N_{23} z \label{jj_theta}
\end{align}
\end{subequations}
where $K\equiv K_{2,3}$, $U\equiv U_{2,3}$ and $N_{23}\equiv N_2+N_3$. 
Following the standard bosonic Josephson junction approach~\cite{SmerziPRL1997,RaghavanPRA1999}, we find that, in the regime of a small population-phase imbalance, Eqs.~\eqref{eq:jj} predict harmonic oscillations of $z(t)$ and $\Delta \theta(t)$ with frequency~\cite{supp}
\begin{equation}\label{eq:omegaJ}
    \omega_J = \sqrt{2K \frac{\alpha}{\alpha-1} U N_{23}} \,.
\end{equation}
Furthermore, Equations~\eqref{eq:jj} predict a MQST regime characterized by oscillations of $z$ around a non-zero value and an approximately linear increase of the phase over the full interval $[0,2\pi]$. 
The transition from SAO to MQST is obtained when the initial population imbalance $z_0$ is larger than the critical value~\cite{supp} 
\begin{equation}\label{eq:zc}
    z_c=\sqrt{\frac{8K}{UN_{23}}\frac{\alpha}{\alpha-1}} \,.
\end{equation}
Equations~(\ref{eq:jj}), (\ref{eq:omegaJ}) and (\ref{eq:zc}) are analogous to those describing the population-phase dynamics in the standard two-mode bosonic Josephson junction~\cite{SmerziPRL1997, RaghavanPRA1999} with the key difference that the lateral clusters renormalize the tunneling coefficient through the parameter $\alpha$ defined in Eq.~(\ref{eq:HpCurrents}).

From the theoretical model Eq.~(\ref{eq:modeljmode}), we can write $\dot{N}_j$ as a function of $K$ and $K' = K_{12} = K_{34}$~\cite{supp}.
We find
\begin{equation}\label{eq:alpha_K}
    \alpha=1+\frac{K}{K'} \sqrt{\frac{N}{N'}},
\end{equation}
where $N = N_2(0)=N_3(0)$ and $N'=N_1(0)=N_4(0)$ denote the equilibrium populations of the central and lateral clusters, respectively. 
Notice that $\alpha = 1+J/J'$, where $J \propto K \sqrt{N}$ and $J' \propto K' \sqrt{N'}$ are the currents at the center and between the central and lateral clusters, respectively (see~\cite{supp} for details).  
In a symmetric and homogeneous system, where $K=K'$ and $N=N'$, the proportionality coefficient simplifies to $\alpha=2$. 
Deviations from this value occur when inhomogeneity or finite-size effects introduces imbalances between the central and lateral clusters dynamics, leading to modified tunneling rates or population distributions.
Therefore, $\alpha$ quantifies the anisotropy of the currents due to the finite trapping potential. 
Remarkably, in the supersolid the population reduction in the lateral clusters produced by the trap is spontaneously compensated by adjusting the tunneling rate $K'$, to achieve $\alpha\approx 2$ even in  systems with a small number of clusters. 

\begin{figure}
    \centering
    \includegraphics[width = \columnwidth]{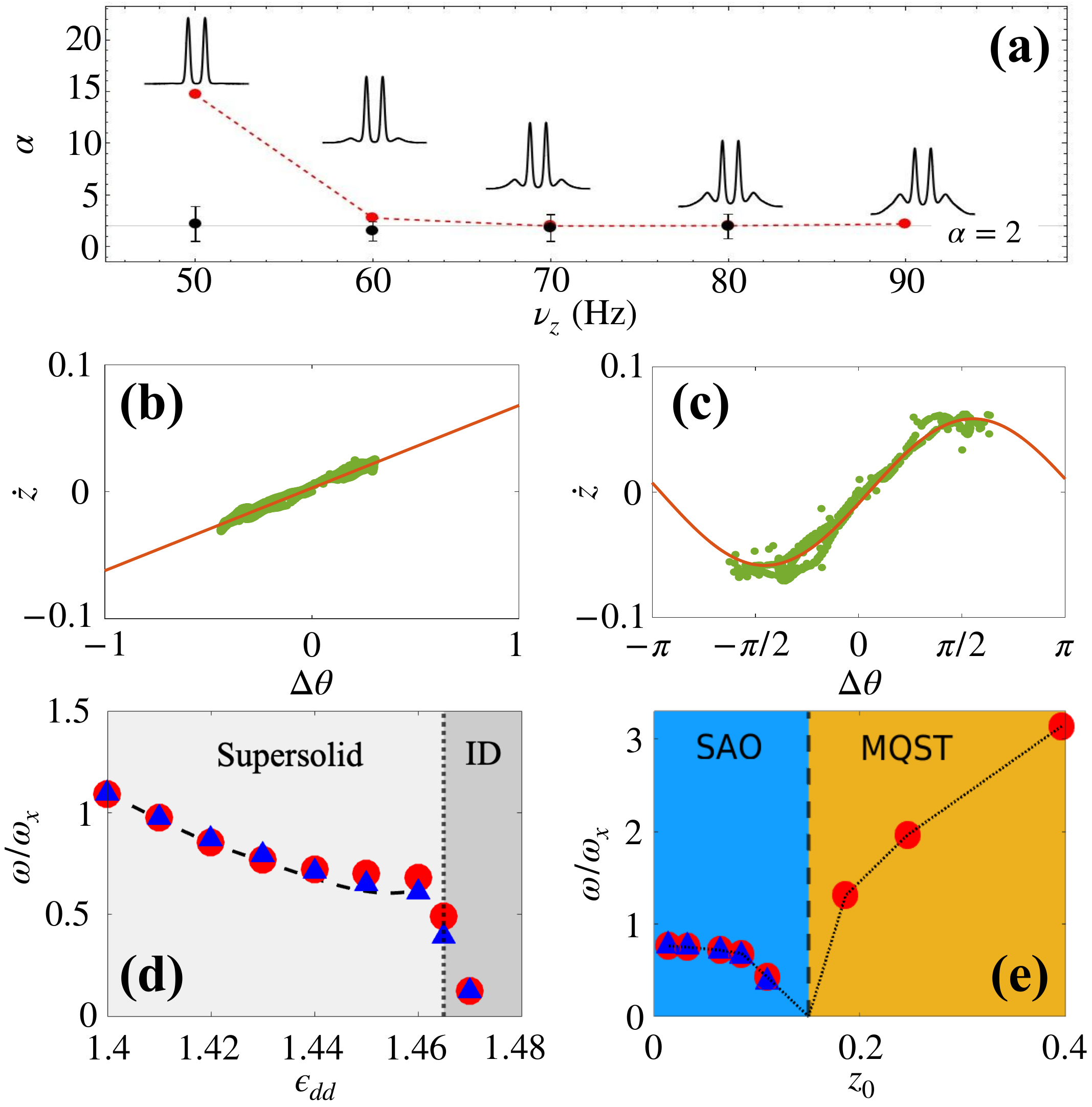}
    \caption{Characterization of the Josephson dynamics.  
    (a) $\alpha$ as a function of $\nu_z$.
    Black dots are the values of $\alpha$ derived from Eq.~\eqref{eq:HpCurrents}, namely from fitting the eGPE dynamics.
    Red dots are obtained from Eq.~\eqref{eq:alpha_K}. 
    The red dotted line is a guide to the eye; the black line highlights the value $\alpha=2$ obtained for a symmetric and homogeneous system.
    Insets are integrated density profiles as a function of $\nu_z$. 
    (b-c) Current-phase relation. Green dots are eGPE results. For small $z_0$ (b), the data are fitted linearly (red line); for larger $z_0$ in the SAO regime (c), the data are well reproduced by a sinusoidal fit (line) using Eq.~\eqref{eq:jj}, yielding the tunneling coefficient $K$.
    (d) Oscillation frequency $\omega$ as a function of $\epsilon_{dd}$ obtained for a small initial imbalance ($z_0 \cong 0.01$).
    Blue triangles and red dots are obtained from an analysis of eGPE data for $z(t)$ and $\Delta\theta(t)$, respectively; the black dashed line is $\omega_J$, Eq.~\eqref{eq:omegaJ}. The vertical dotted line marks the supersolid-to-ID transition.  
    (e) Oscillation frequency $\omega$ as a function of the initial imbalance $z_0$. 
    The vertical black dashed line shows the critical imbalance $z_c$, Eq.~\eqref{eq:zc}, marking the SAO (blue) to MQST (orange) transition. 
    The dotted line is a guide to the eye. Here, $\epsilon_{dd} = 1.43$.
    All frequencies are renormalised by the trap frequency $\omega_x=2\pi\nu_x$. 
   }\label{fig:CharacterizationJosDyn}
\end{figure}

To demonstrate this point, we study the behavior of $\alpha$ as a function of $\nu_z$, comparing the predictions of our model with the dynamical simulations.
Considering Eq.~\eqref{eq:alpha_K}, we calculate $K$ and $K'$ directly from the ground state density profile, exploiting the relation between the Josephson tunneling and the superfluid fraction associated with each junction~\cite{biagioni2023subunity,supp}, and we obtain $\alpha$ predicted by the Josephson model.
In Fig.~\ref{fig:CharacterizationJosDyn}(a), we compare these static values (red dots) with the data extracted from the dynamics (black dots) using Eq.~\eqref{eq:HpCurrents}, which does not depend on the model.
For large $\nu_z$, the system approaches the homogeneous limit where $\alpha=2$. 
Decreasing $\nu_z$, the system transits into the MO regime and Eq.\,\eqref{eq:alpha_K} (red dots) -- which accounts for the enhancement of the inhomogeneity introduced by the finite trapping potential on equilibrium properties -- predicts an increasing value of $\alpha$. 
Around $\nu_z=50$ Hz, the values extracted using Eq.\,\eqref{eq:HpCurrents} (black dots) -- accounting for the numerical dynamics -- deviate from the model.
This discrepancy arises from the coupling of the Josephson oscillations with the Goldstone mode at low $\nu_z$.
%
%
When the lateral clusters are sufficiently small, the initial shift of the center of mass, used to trigger the Josephson mode, excites a 
low-energy Goldstone mode: a current counterbalances the motion of the clusters, reducing the energy cost with respect to $\hbar\nu_x$. 
This mode couples with the Josephson one, resulting in the loss of current parity and, consequently, the inability to define a single $\alpha$. 
%
%
We finally point out that the above equations and discussion can be extended to a generic linear system with an arbitrary number of clusters.
%

%
We now focus on the region of parameters where the 
Josephson dynamics is decoupled from the Goldstone mode. 
In this regime, we compare the predictions of Eqs.~(\ref{eq:jj}), (\ref{eq:omegaJ}) and (\ref{eq:zc}) with the numerical dynamics.
In particular, the dotted lines in Fig.~\ref{fig:Josephsondynamics} are obtained by the direct integration of Eq.~(\ref{eq:modeljmode}), depending on the parameters $U_j$, $E_j$ and $K_j$. 
To obtain the interaction and tunneling parameters, we write the current-phase relation Eqs.~(\ref{eq:jj}) in the linear regime: $\Dot{z} = 2K\frac{\alpha}{\alpha-1}\Delta\theta$ and $\Dot{\Delta\theta} = -UN_{23}z$, extracting $K$ and $UN_{23}$ by a linear fitting using $z(t)$ and $\Delta\theta(t)$ from the eGPE dynamics~\cite{supp}. 
An example of the linear current-phase relation from the eGPE data is shown in Fig.~\ref{fig:CharacterizationJosDyn}(b) (green dots) along with its linear fit (red line). 
The same analysis of other specific combinations of Eqs.~(\ref{eq:modeljmode}) allows for the extraction of the remaining parameters $K_{j,j+1}$ and $U_{j}$~\cite{nota2}. %
For higher values of $z_0$, see Fig.~\ref{fig:CharacterizationJosDyn}(c), the current-phase relation becomes non-linear, but still $K$ can be extracted from a sinusoidal fit, obtaining the same value of the linear regime. 
It can be seen that the values of the tunneling extracted from the ground state density profile agree with those obtained from the current-phase relations.

From a sinusoidal fit of the population and phase oscillations in the eGPE we extract the oscillation frequency $\omega$ as a function of $\epsilon_{dd}$, restricting to the small-amplitude regime. 
Results are shown in Fig.~\ref{fig:CharacterizationJosDyn}(d) by blue triangles and red circles for $z$ and $\Delta\theta$, respectively.  
We observe a general decreasing trend~\cite{biagioni2023subunity} due to the increasing depth of density modulation with $\epsilon_{dd}$ that we associate to the decrease of the effective tunneling or, equivalently, a reduction of the superfluid fraction~\cite{leggett_1970}. 
A sudden drop in the Josephson oscillation frequency witnesses the transition from SAO to MQST regimes. 
In Fig.~\ref{fig:CharacterizationJosDyn}(e), we plot the oscillation frequency $\omega$ as a function of the initial population imbalance $z_0$ for $\epsilon_{dd}=1.43$. 
In the SAO regime (blue region in the figure), $\omega$ remains almost constant but drops when approaching the MQST regime (orange region). 
The vertical black dashed line is the theoretical value $z_c$, given by Eq.~\eqref{eq:zc}. 

\begin{figure}
    \centering
    \includegraphics[width = 1\linewidth]{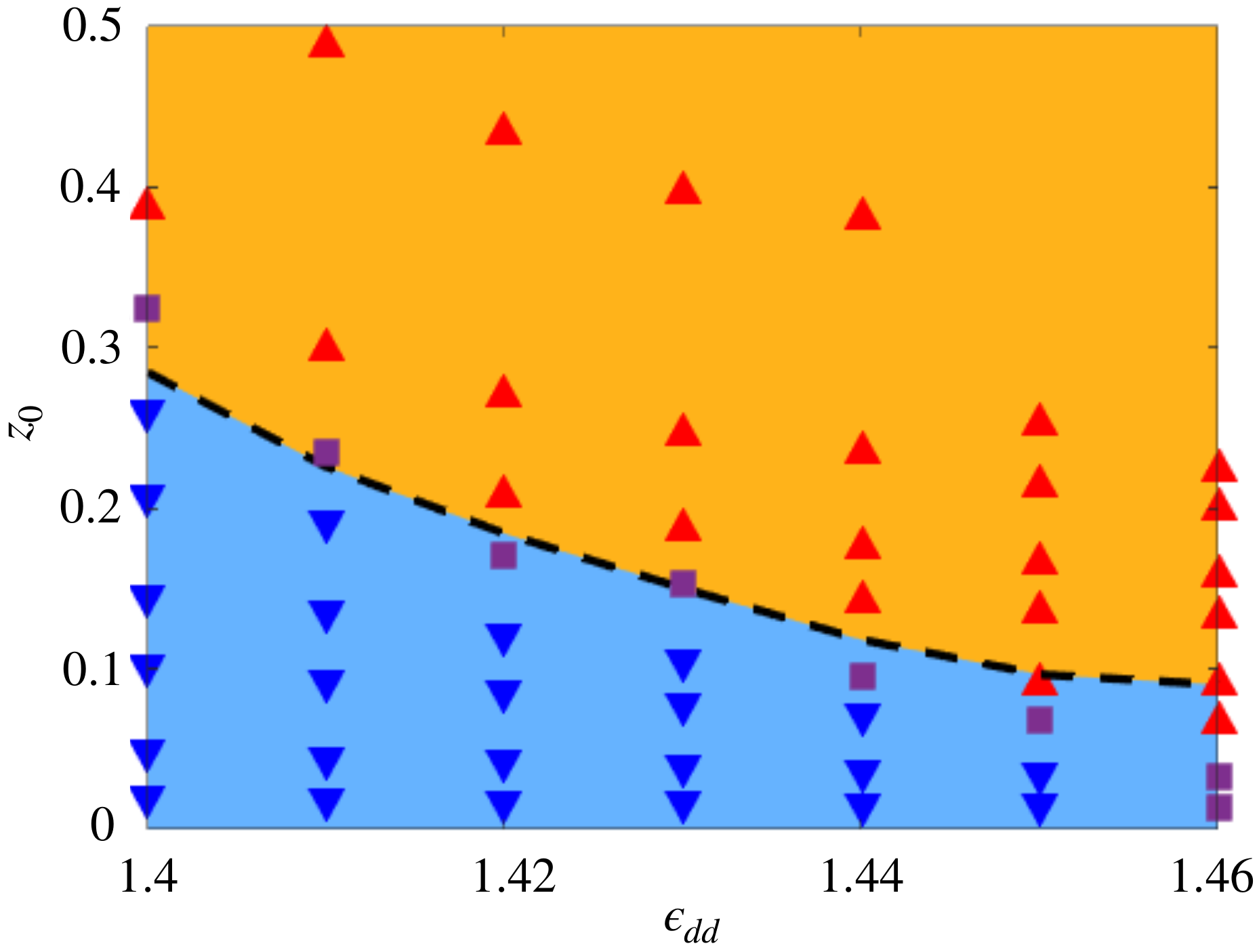}
    \caption{Phase diagram showing the transition from SAO, for $z<z_c$ (blue region), to MQST, for $z>z_c$ (orange region), as a function of the interaction strength $\epsilon_{dd}$ and $z_0$. 
    Each triangle corresponds to a numerical result of the eGPE: blue downward triangles corresponds to $\bar{z}=0$, red upward triangles correspond to $\bar{z} \neq 0$, while purple squares correspond to a transitory regimes. 
    The black dashed line is Eq.~(\ref{eq:zc}). 
    Here, $\nu_z = 80$ Hz.}
    \label{fig:Zc}
\end{figure}

The boundary between the SAO and MQST is presented, as a function of the initial imbalance $z_0$ and the interaction $\epsilon_{dd}$, in Fig.~\ref{fig:Zc}. 
In the SAO regime (blue region) we numerically observe oscillations around $\bar{z} \equiv \tfrac{1}{T}\int_{0}^T dt z(t)=0$, with $T$ encompassing a few oscillations (blue downward triangles). 
The MQST regime (orange region) occurs when $\bar{z}\neq 0$ (red upward triangles). 
Simulations corresponding to purple squares show a mixed chaotic behavior.
For instance, the dynamics may be initially self-trapped and later both $z$ and $\Delta\theta$ oscillate around $z=0$, as in the SAO regime. 
The black dashed line is the theoretical critical value $z_c$ in Eq.~\eqref{eq:zc} and agrees well with the numerical results. 

\textit{Conclusions ---}
We have shown that the Josephson effect can appear in supersolids with dynamical regimes analogous to those observed in Bose gases coupled by external potential barriers, at least in regimes in which one can neglect the coupling to the Goldstone mode. 
In particular, a supersolid can be described as a self-induced Josephson junction array, with a model that takes into account the inhomogeneity due to the trapping potential, which captures quantitatively the various phenomena.
In the past, analytical models of Josephson junctions realized with potential barriers in BECs have allowed the study of thermal~\cite{GatiPRL2006} and quantum fluctuations~\cite{MilburnPRA1997, EsteveNATURE2008}, the latter having interesting applications in entanglement-enhanced interferometry~\cite{PezzePRA2006, BerradaNATCOMM2013, PezzeRMP2018}.
Our study will facilitate the investigation of such effects in supersolids, eventually extending the analysis and modeling to include the Goldstone mode.\\

{\it Note added in Proof.} 
A related preprint~\cite{AlanaARXIV} appeared quite recently, investigating the Josephson dynamics in a triangular dipolar supersolid. \\

\begin{acknowledgments}
{\it Acknowledgments.}
This work has been supported by QuantERA project SQUEIS (Squeezing enhanced inertial sensing), funded by the European Union’s Horizon Europe Program and the Agence Nationale de la Recherche (ANR-22-QUA2-0006); 
the Horizon Europe Program HORIZONCL4-2022-QUANTUM-02-SGA via the project 101113690 (PASQuanS2.1);
the European Union (European Research Council, SUPERSOLIDS, Grant No. 101055319) and by the QuantERA program, project MAQS, under Grant No. 101017733, with funding organization Consiglio Nazionale delle Ricerche. We acknowledge support from the European Union, NextGenerationEU, for the “Integrated Infrastructure initiative in Photonics and Quantum Sciences” (I-PHOQS Grant No. IR0000016, ID Grant No. D2B8D520, and CUP Grant No. B53C22001750006) and for PNRR MUR Project NQSTI-PNNR-PE4 No. PE0000023.  
\end{acknowledgments}

\renewcommand{\theequation}{S\arabic{equation}}
\setcounter{equation}{0}

\clearpage

\onecolumngrid
\section{Supplementary Material}
\twocolumngrid

\subsection{Theoretical derivation of Josephson equations}
Consider the four coupled Eqs.~(1) that describe the evolution of the four localized wavefunctions \cite{SmerziPRL1997}. Expressing them as $\psi_j(t)=\sqrt{N_j(t)}e^{-i\theta_j(t)}$, the evolution of the four phases can be written as
\begin{equation}
\begin{split}
    \dot{\theta_1} = - (E_1 + N_1 U_1) \\
    \dot{\theta_2} = - (E_2 + N_2 U_2) \\
    \dot{\theta_3} = - (E_3 + N_3 U_3) \\
    \dot{\theta_4} = - (E_4 + N_4 U_4) \,,
\end{split}    
\end{equation}
where we supposed $K\ll N U$. The phase difference between the two central clusters $\Delta\theta\equiv \Delta\theta_{23}$ evolves as the Eq.~(3b)
\begin{equation}
\begin{split}
    \dot{\Delta \theta} &= \dot{\theta_3} - \dot{\theta_2} = - (E_3 + N_1 U_3) + (E_2 + N_1 U_2) = \\
    &= - U (N_3 - N_2) = - U (N_2 + N_3) z \,,
\end{split}
\end{equation}
where we have used the symmetry of the system $E_3=E_2$ and $U_3=U_2$. 
The populations evolution of the four clusters are given by
\begin{equation}
\begin{split}
    \dot{N_1} &= -2K' \sqrt{N_1 N_2} \sin{\Delta \theta_{12}} \\
    \dot{N_2} &= 2K' \sqrt{N_1 N_2} \sin{\Delta \theta_{12}} - 2K \sqrt{N_2 N_3} \sin{\Delta \theta_{23}} \\
    \dot{N_3} &= 2K  \sqrt{N_2 N_3} \sin{\Delta \theta_{23}} - 2K' \sqrt{N_3 N_4} \sin{\Delta \theta_{34}} \\
    \dot{N_4} &= 2K' \sqrt{N_3 N_4} \sin{\Delta \theta_{34}} \,, 
\end{split}
\end{equation}
where $K=K_{23}$ and $K'=K_{12}=K_{34}$. 
To decouple these equations and write the Josephson equations for the central cluster $2-3$, we use Eq.~(2)
\begin{equation}
\begin{split}
    \dot{N_3} - \dot{N_2} &= 4K \sqrt{N_2 N_3} \sin{\Delta \theta_{23}} \\
     - 2K' &\sqrt{N_3 N_4} \sin{\Delta \theta_{34}} - 2K' \sqrt{N_1 N_2} \sin{\Delta \theta_{12}} = \\
    &= 4K \sqrt{N_2 N_3} \sin{\Delta \theta_{23}} + \dot{N_1} - \dot{N_4} = \\
    &= 4K \sqrt{N_2 N_3} \sin{\Delta \theta_{23}} + (\dot{N_3} - \dot{N_2}) /\alpha \,,
\end{split}    
\end{equation}
so that
\begin{equation}
    \dot{N_3} - \dot{N_2} = 4K \frac{\alpha}{\alpha-1} \sqrt{N_2 N_3} \sin{\Delta \theta_{23}} \,.
\end{equation}
Assuming, as a first approximation, that the sum of the two central clusters is constant, the left-hand side simplifies to $\dot{z}$. Writing
\begin{equation}
    \frac{1 \pm z}{2} = \frac{1}{2} \left( 1 \pm \frac{N_3-N_2}{N_2+N_3} \right) = \frac{N_{2,3}}{N_2+N_3} \,,
\end{equation}
we find $\sqrt{N_2 N_3}=\sqrt{1-z^2}/2$. Substituting, we obtain
\begin{equation}
    \dot{z} = 4K \frac{\alpha}{\alpha-1} \sqrt{1-z^2}/2 \sin{\Delta \theta} \,,
\end{equation}
which corresponds to Eq.~(3a). 

\subsection{Josephson frequency and critical population imbalance for MQST}
Starting from Eqs.~(3) and following the approach for a standard bosonic Josephson junction, we can derive the Josephson frequency $\omega_J$ for small oscillations, thus in the SAO regime. We write the second derivative of $z$ with respect to time approximating the equations for small values of both $z$ and $\Delta\theta$
\begin{equation}
    \Ddot{z} \cong 2K \frac{\alpha}{\alpha-1} \dot{\Delta\theta} = - 2K \frac{\alpha}{\alpha-1} U N_{23} z \,.
\end{equation}
This is exactly the equation of a simple harmonic oscillator with frequency given by Eq.~(4). 

To derive the critical initial populayion imbalance $z_c$ necessary for the MQST regime, we need to pass from the Hamiltonian $H$. Considering $z$ and $\Delta\theta$ as two conjugate variables, thus $\dot{z}=-\frac{\partial H}{\partial \Delta\theta}=2K\alpha/(\alpha-1)\sin{\Delta\theta}$ and $\dot{\Delta\theta}=\frac{\partial H}{\partial z}=-UN_{23}z$, so that
\begin{equation}
    H(z,\Delta\theta) = 2K \frac{\alpha}{\alpha-1} \cos{\Delta\theta} - U N_{23} \frac{z^2}{2} \,.
\end{equation}
For small values of $z$ and $\Delta\theta$ the phase-space is like the simple pendulum one, until the so-called separatrix line, over which the system enters the MQST regime. The value of $z_c$ can be found imposing $H(0,\pi)=H(z_c,0)$
\begin{equation}
\begin{split}
    -2K \frac{\alpha}{\alpha-1} &= 2K \frac{\alpha}{\alpha-1} - U N_{23} \frac{z^2}{2} \\
    -4K \frac{\alpha}{\alpha-1} &= - U N_{23} \frac{z^2}{2} \\
    z_c^2 &= \frac{8K\frac{\alpha}{\alpha-1}}{N_{23}U} \,,    
\end{split}
\end{equation}
that corresponds to Eq.~(5).

\subsection{Expression of $\alpha$ from density state}
The parameter $\alpha$ given in Eq.~(6) can be derived from its definition in Eq.~(2). Assuming small phase and population oscillations, we express $\alpha$ as
\begin{equation}
\begin{split}
    &\alpha = \frac{\dot{N_3}-\dot{N_2}}{\dot{N_1}-\dot{N_4}} = \\
    &= (4K \sqrt{N_2^0 N_3^0} \Delta \theta_{23} - 2K' \sqrt{N_3^0 N_4^0} \Delta \theta_{34} - 2K' \sqrt{N_1^0 N_2^0} \Delta \theta_{12}) / \\
    &(- 2K' \sqrt{N_1^0 N_2^0} \Delta \theta_{12} - 2K' \sqrt{N_3^0 N_4^0} \Delta \theta_{34}) = \\
    &= 1 + 4K \sqrt{N_2^0 N_3^0} \Delta \theta_{23} / (- 2K' \sqrt{N_1^0 N_2^0} \Delta \theta_{12} - 2K' \sqrt{N_3^0 N_4^0} \Delta \theta_{34}) \,.
\end{split}    
\end{equation}
For a symmetric system where $N_1^0=N_4^0$ and $N_2^0=N_3^0$, and with alternating phase differences, like a Josephson dynamics ($\Delta\theta_{12}=-\Delta\theta_{23}=\Delta\theta_{34}$), the expression simplifies to
\begin{equation}
    \alpha = 1 + \frac{K}{K'} \sqrt{\frac{N_2^0}{N_1^0}} \,.
\end{equation}
The tunnelings $K$ and $K'$ need to be determined from dynamical data fits. 
However, an alternative expression for $\alpha$ can be derived entirely from the density profile of the equilibrium state. Using the result of \cite{biagioni2023subunity}, the tunneling can be expressed as a function of the superfluid fraction $K_{ij} = f_{s,ij} \frac{\hbar^2}{2m d^2}$, where $d$ is the supersolid spacing and $f_{s,ij}$ is the superfluid fraction calculated using Leggett's definition \cite{leggett_1970} within the unit cell between the two clusters $i$ and $j$.
Substituting this into the expression for $\alpha$, we obtain
\begin{equation}
    \alpha = 1 + \frac{f_{s,23}}{f_{s,12}} \sqrt{\frac{N_2^0}{N_1^0}} \,.
\end{equation}
This formulation allows $\alpha$ to be determined directly from the density profile, providing an alternative method to dynamical data fits.

\subsection{Numerical methods}
We study the Josephson dynamics of the supersolid with the numerical integration of the extended Gross-Pitaevskii equation (eGPE)
\begin{align}\label{eGPE}
    i \frac{\partial \psi(\mathbf{r}, t)}{\partial t} &= \biggl[ -\frac{\nabla^2}{2m} + V_t(\mathbf{r}) + g|\psi(\mathbf{r},t)|^2 \\ &+ \int d\mathbf{r'} V_{dd}(\mathbf{r}-\mathbf{r'}) |\psi(\mathbf{r'},t)|^2
    + \gamma(\epsilon_{dd})|\psi(\mathbf{r},t)|^3 \biggl] \psi(\mathbf{r},t) \,.
\end{align}
This equation has been extensively used in theoretical studies of a dipolar-gas supersolid \cite{Roccuzzo2019,LimaPelster2011,LimaPelster2012,WchtlerPRA2016,Petrov2015_eGPE,Alaa2022,RoccuzzoPRL2020}, finding excellent agreement with experimental findings \cite{TanziPRL2019,SS_Ferlaino,SS_Pfau,BiagioniPRX2022,biagioni2023subunity,Tanzi2021}. 
In \eqref{eGPE}, $V_t(\mathbf{r}) = \frac{1}{2}m(\omega_x^2 x^2+\omega_y^2 y^2 + \omega_z^2 z^2)$ is an harmonic trapping potential, $g=\frac{4\pi\hbar^2 a_s}{m}$ is the contact interaction parameter and $V_{dd}({\mathbf{r}}) = \frac{C_{dd}}{4\pi}\frac{1-3\cos^2\theta}{r^3}$ is the dipolar interaction. In addition to the standard Gross-Pitaevskii equation, \eqref{eGPE} includes the LHY term calculated within the local density approximation for a dipolar system \cite{LimaPelster2011,LimaPelster2012} and has the form $\gamma(\epsilon_{dd}) = \frac{32}{3\sqrt{\pi}} g a_s^{3/2} F(\epsilon_{dd})$ where $F(\epsilon_{dd}) = \frac{1}{2}\int_0^\pi d\theta \sin\theta[1+\epsilon_{dd}(3\cos^2\theta - 1)^{5/2}]$ and $\epsilon_{dd} = a_{dd}/a_s$. 

We found the ground states of our system by evolving the eGPE \eqref{eGPE} in imaginary time using the fourth-order Runge-Kutta technique implemented by a dynamic time step adjustment method \cite{ITP_dtLS_Lehtovaara2007}. 
We first had to work out the dipolar term so that we could reduce the eGPE to an ordinary differential equation. To do this, we have noticed that the dipolar term $\Phi_{dd}(\textbf{r},t)$ has the form of a convolution, so that it is smart to evaluate it in the Fourier space. Indeed, the convolution theorem states that the Fourier transform of a convolution of two functions is the product of the Fourier transforms of the two functions $\mathcal{F}[f*g] = \mathcal{F}[f] \cdot \mathcal{F}[g]$, where the convolution is defined as $f*g = \int d\mathbf{r} f(\mathbf{r}) g(\mathbf{r} - \mathbf{r'})$. The dipolar term can therefore be rewritten as: 
\begin{equation}
    \Phi_{dd} (\mathbf{r}, t) = \mathcal{F}^{-1} \bigl[ \mathcal{F}[V_{dd}](\mathbf{k}) \cdot  \mathcal{F}[n](\mathbf{k}) \bigl]
\end{equation}
where $\mathcal{F}^{-1}$ is the inverse Fourier transform and we have called $n$ the density $|\phi(\mathbf{r}, t)|^2$. The expression of $\mathcal{F}[V_{dd}]$ can be calculated analitically and takes the form: 
\begin{equation}\label{FT_dipolare_tot}
    \Tilde{V}_{dd}(\mathbf{k}) = \frac{C_{dd}}{3} (3 \cos^2 \beta -1) 
\end{equation}
where $\beta$ is the angle between the dipoles and the $\textbf{k}$ vector. So, we need to perform a numerical Fourier transform of the density using the Fast Fourier Transform (FFT) algorithm, multiply it by the Fourier transform of the dipolar interaction \eqref{FT_dipolare_tot}, then calculate the inverse Fourier transform of the product with the inverse FFT. 

Having reduced the eGPE to an ordinary differential equation, before applying the integration method we must discretize the space, defining a spatial grid, i.e. a vector $x$ starting from $-L_x$ and ending at $L_x$ with a step equal to $dx$, and the same for $y$ and $z$. The steps are defined using the number of them $N_i$ in the corresponding direction $ dx_i = \frac{2L_i}{N_i-1} \qquad i=x, y, z$ with $N_x = 128$, $N_y = N_z = 64$. Notice that we have chosen numbers of points that are powers of two because we have to use the FFT. Another necessary thing to do for FFT is to define another grid, but in the momentum space $k_i = - \frac{N_i}{2} dk_i : dk_i : \frac{N_i}{2} dk_i$ with the momentum step defined with respect to the maximum space length $ dk_i = \frac{\pi}{2 L_i}$, so that the $k_i$ vectors are the same length of the corresponding $x_i$ vectors. Now, we can discretize the other terms of the eGPE. 

The first term is the kinetic one, so, unless a numerical pre-factor, it essentially is the 3D laplacian that can be written on the discrete grid using finite differences at different orders. We have used the finite difference 5-point formula $\frac{\partial^2 f}{\partial x^2} = (-f(x-2dx) + 16 f(x-dx) - 30 f(x) + 16 f(x+dx) - f(x+2dx))/(12 dx^2)$
in all three directions. 

The external harmonic potential is $V(x, y, z) = \frac{1}{2}m (\omega_x^2 x^2 + \omega_y^2 y^2 + \omega_z^2 z^2)$ and contains the mass $m\cong 161.9268u$ of the $^{162}Dy$, where $u$ is the atomic mass $u=1.67\cdot 10^{-27} Kg$ and the frequencies are the ones mentioned in the main text. 

In order to observe Josephson oscillations, it is necessary to introduce an initial non-zero population imbalance in the two central clusters by configuring them in an out-of-equilibrium state. This is achieved in our simulations by determining the ground state of the system with the addition of an external sinusoidal potential $V_{sin}(x) = -\gamma \sin{\biggl(\frac{\pi}{d} \ x \biggl)}$, where $d$ is the distance between the two central clusters, such that one maximum of the potential corresponds to the position of one cluster, while the subsequent minimum to the other cluster. The amplitude of the potential is of the same order as the harmonic potential $\gamma \sim \hbar \omega_x$ and can be adjusted to produce different initial imbalances $z_0$ in the system. This sinusoidal potential can be generated experimentally using an additional optical lattice created by a standing laser beam.

Even more accurate than a 4th order Runge Kutta method, it is the same method but with an adaptive stepsize algorithm. In this work we have used the one proposed by Lehtovaara \textit{et al.} \cite{ITP_dtLS_Lehtovaara2007}. It consists in using two different time steps for the integration and the algorithm makes the code choosing the best time step and adjusts it dynamically. In particular, the selection criteria is based on the time step giving the smaller energy. The code makes this choice every iteration, but it is allowed only to keep the same $dt$ or decreasing it, never increasing it. So, the initial time step need to be chosen as large as possible, so as not to make it unstable. In particular, we can use the condition stability for a 3D diffusion-type equation $dt \leq \frac{dx^2+dy^2+dz^2}{2D}$, where $D$ is the diffusion coefficient that in our case is $D=\frac{\hbar^2}{2m}$. 

These iterations need a convergence criterion that tells them when they need to stop. To fix this criterion, we use the total energy. We have already included the energy calculation in the dynamic time step adjustment method as: 
\begin{equation}
    E_{tot}^n = E_{kin}^n + E_{h.o.}^n + E_{con}^n + E_{dd}^n + E_{LHY}^n
\end{equation}
where: 
\begin{equation}
\begin{split}
    E_{kin}^n &= \bra{\phi_n} \hat{K} \ket{\phi_n} = \int d\mathbf{r} \frac{\hbar^2}{2m} |\nabla n(\mathbf{r})|^2\\ 
    E_{h.o.}^n &= \bra{\phi_n} \hat{V}_{h.o.} \ket{\phi_n} = \int d\mathbf{r} \  \frac{1}{2} m (\omega_x^2 x^2 + \omega_y^2 y^2 + \omega_z^2 z^2) n(\mathbf{r})\\
    E_{con}^n &= \bra{\phi_n} \hat{V}_{contact} \ket{\phi_n} = \frac{1}{2} g \int d\mathbf{r}  n(\mathbf{r})\\
    E_{dd}^n &= \bra{\phi_n} \hat{V}_{dd} \ket{\phi_n} = \frac{1}{2} \int d\mathbf{r} \int d\mathbf{r'} V_{dd}(|\mathbf{r}-\mathbf{r'}|) n(\mathbf{r'}) n(\mathbf{r}) \\
    E_{LHY}^n &= \bra{\phi_n} \hat{V}_{LHY} \ket{\phi_n} = \frac{2}{5} \gamma(\epsilon_{dd}) \int d\mathbf{r} n(\mathbf{r})^{5/2} 
\end{split}
\end{equation}
and $n(\mathbf{r}) = |\phi(\mathbf{r})|^2$ is the density calculated at the n-th step. Every iteration, this energy is compared to the one calculated at the step before: 
\begin{equation}
    \frac{E_{tot}^{n-1} - E_{tot}^{n}}{E_{tot}^{n-1}}
\end{equation}
Iterations keep going until the check variable reaches a value below a certain threshold, which we have set to $10^{-8}$.

\subsection{Ground state}
\begin{figure}[bh]
    \centering
    \includegraphics[width = 0.49\columnwidth]{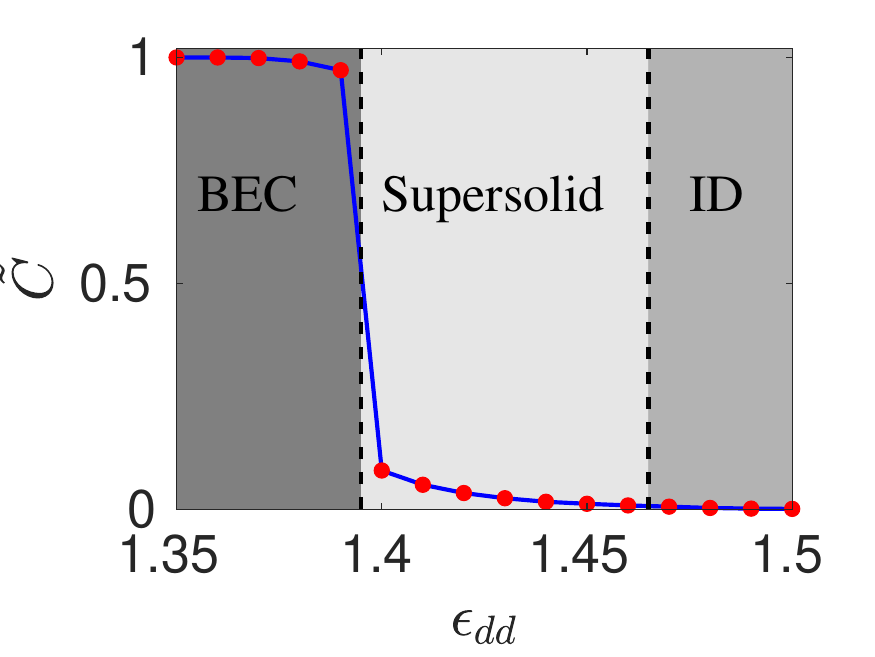}
    \includegraphics[width = 0.49\columnwidth]{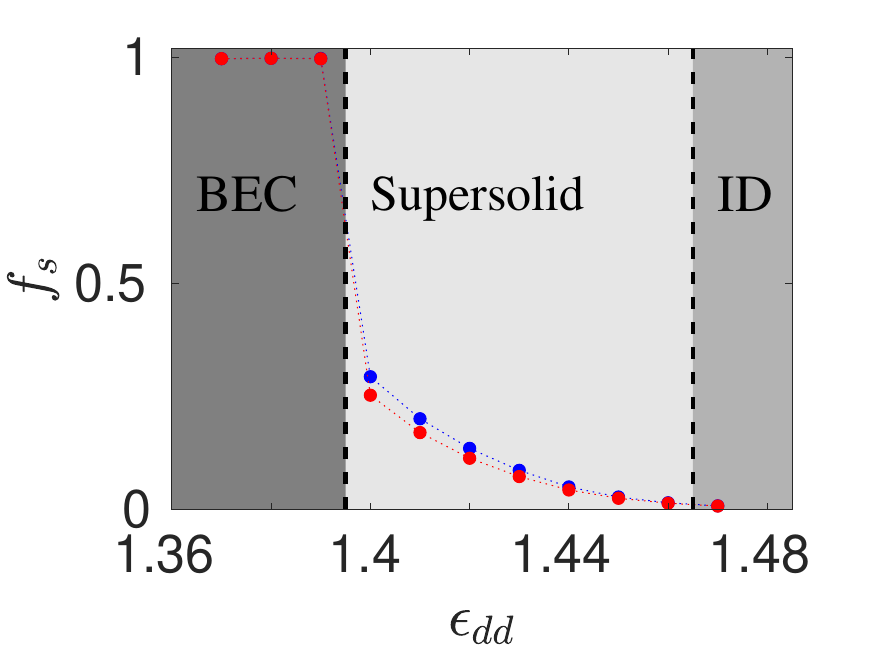}
    \caption{Strength of the density modulation $\Tilde{C}$ and the upper (blue points) and lower bound (red points) of the superfluid fraction $f_s$ as a function of the interaction parameter $\epsilon_{dd}$. They both assume the value of 1 in the BEC regime and, as the supersolid regime is entered, they decrease and ultimately attain a value of zero in the IDs regime, where there is no longer any coupling between clusters. 
    }
    \label{fig:Ctilde}
\end{figure}

Equilibrium states have been determined for various values of the interaction $\epsilon_{dd}$. In the BEC regime, the density exhibits no modulation, but as the interaction is increased, a modulation in the density is observed to emerge. The strength of this modulation depends on the interaction and can be estimated by the dimensionless modulation contrast $\tilde{C}$, defined as:
\begin{equation}\label{Ctilde}
    \tilde{C} = 1 - \frac{n_{max} - n_{min}}{n_{max} + n_{min}}
\end{equation}
where $n_{min}$ and $n_{max}$ are the minimum between the two central clusters and the maximum values of the 1d integrated equilibrium density $n(x)$, respectively (see Fig.\,\ref{fig:Ctilde}). 

In order to further characterize the transition to the supersolid regime, the superfluid fraction is also considered. The upper and lower bounds of the superfluid fraction, as defined by Leggett \cite{leggett_1970,ZapataPRA1998}, are
\begin{equation}\label{fs}
     \int \frac{dy dz}{\int_{UC} \frac{dx}{\Bar{n}(x,y,z)}} \leq fs \leq \biggl(\int_{UC} \frac{dx}{\int dy dz \ \Bar{n}(x,y,z)} \biggl)^{-1}
\end{equation}
where the integral over x is performed over the unitary cell UC, which is the interval between the two maxima of the central clusters and $\Bar{n}$ is the density normalized over it. In conclusion, the analysis of the equilibrium state reveals the emergence of a density modulation in the BEC regime as the interaction is increased, and the superfluid fraction serves as a useful tool in characterizing the transition to the supersolid regime.

\begin{thebibliography}{100}

\bibitem{JosephsonPL1962}
B.D. Josephson, Possible new effects in superconductive tunneling. 
{\it Physics Letters} {\bf 1}, 251 (1962).

\bibitem{BaroneBOOK}
A. Barone and G. Paternò {\it Physics and Applications of the Josephson Effect} (John Wiley \& Sons, London, 1982).

\bibitem{ReviewPackardSF}
J. C. Davis and R. E. Packard, Superfluid $^{3}$He Josephson
weak links. 
{\it Rev. Mod. Phys.} {\bf 74}, 741 (2002). 

\bibitem{CataliottiSCIENCE2001}
F. S. Cataliotti, et al., 
Josephson Junction Arrays with Bose-
Einstein Condensates. 
{\it Science} {\bf 293}, 843 (2001).

\bibitem{AlbiezPRL2005}
M. Albiez, et al., 
Direct Observation of Tunneling and Nonlinear Self-Trapping in a Single Bosonic Josephson Junction. 
{\it Phys. Rev. Lett.} {\bf 95}, 010402 (2005). 

\bibitem{ValtolinaSCIENCE2015}
G. Valtolina, et al., 
Josephson effect in fermionic superfluids across the BEC-BCS crossover. 
{\it Science} {\bf 350}, 9203 (2015).

\bibitem{SpagnolliPRL2017} 
G. Spagnolli, et al., 
Crossing Over from Attractive to Repulsive Interactions in a Tunneling Bosonic Josephson Junction.
{\it Phys. Rev. Lett.} {\bf 118} 230403 (2017).

\bibitem{KwonSCIENCE2020}
W. J. Kwon, et al., 
Strongly correlated superfluid order parameters from dc Josephson supercurrents. 
{\it Science} {\bf 369} 84, (2020).

\bibitem{SmerziPRL1997}
A. Smerzi, S. Fantoni, S. Giovanazzi, and S. R. Shenoy, Quantum Coherent Atomic Tunneling between Two Trapped Bose-Einstein Condensates. 
{\it Phys. Rev. Lett.} {\bf 79}, 4950 (1997). 

\bibitem{RaghavanPRA1999}
S. Raghavan, A. Smerzi, S. Fantoni, and S. R. Shenoy, Coherent oscillations between two weakly coupled Bose-Einstein condensates: Josephson effects, $\pi$ oscillations, and macroscopic quantum self-trapping. 
{\it Phys. Rev. A} {\bf 59}, 620 (1999).

\bibitem{GrossPRL1957}
E. P. Gross, 
Unified Theory of Interacting Bosons.
{\it Phys. Rev.} {\bf 106}, 161 (1957)

\bibitem{AndreevJETP1969}
A. F. Andreev and I. M. Lifshitz, 
Quantum theory of defects in crystals. 
{\it Sov. Phys. JETP} {\bf 29}, 1107 (1969).

\bibitem{ChesterPRA1970}
G. V. Chester,
Speculations on Bose-Einstein Condensation and Quantum Crystals.
{\it Phys. Rev. A} {\bf 2}, 256 (1970).

\bibitem{leggett_1970}
A. J. Leggett, Can a Solid Be ``Superfluid''?
{\it Phys. Rev. Lett.} {\bf 25} 1543 (1970).

\bibitem{TanziPRL2019}
L. Tanzi, et al., Observation of a dipolar quantum gas with metastable supersolid properties. 
{\it Phys. Rev. Lett.} {\bf 122}, 130405 (2019).

\bibitem{SS_Ferlaino}
L. Chomaz, et al., Long-Lived and Transient Supersolid Behaviors in Dipolar Quantum Gases. 
{\it Phys. Rev. X} {\bf 9}, 021012 (2019). 

\bibitem{SS_Pfau}
F. B\"ottcher, et al., 
Transient Supersolid Properties in an Array of Dipolar Quantum Droplets. 
{\it Phys. Rev. X} {\bf 9}, 011051 (2019). 

\bibitem{KunimiPRB2011}
M. Kunimi, Y. Nagai, and Y. Kato,
Josephson effects in one-dimensional supersolids.
{\it Phys. Rev. B} {\bf 84}, 094521 (2011).

\bibitem{AbadEPL2011}
M. Abad, M. Guilleumas, R. Mayol, M. Pi, and D. M. Jezek, 
A dipolar self-induced bosonic Josephson
junction. 
{\it Europhys. Lett.} {\bf 94} 10004 (2011).

\bibitem{Ilzhfer2021}
P. Ilzh\"ofer, et al., 
Phase coherence in out-of-equilibrium supersolid states of ultracold dipolar atoms. 
{\it Nature Physics} {\bf 17}, 356 (2021).

\bibitem{BuhlerPRR2023}
C. B\"uhler, T. Ilg, and H. P. B\"uchler,
Quantum fluctuations in one-dimensional supersolids.
{\it Phys. Rev. Research} {\bf 5}, 033092 (2023)

\bibitem{PlattARXIV}
L. M. Platt, D. Baillie, P. B. Blakie,
Supersolid spectroscopy.
arXiv:2412.15552. 

\bibitem{biagioni2023subunity}
G. Biagioni, N. Antolini, B. Donelli, L. Pezzè, A. Smerzi, M. Fattori, A. Fioretti, C. Gabbanini, M. Inguscio, L. Tanzi, and G. Modugno, Measurement of the superfluid fraction of a supersolid by Josephson effect. 
{\it Nature} {\bf 629}, 773-777 (2024).

\bibitem{supp}
See Supplementary Material for details of the analytical model and numerical methods. 
The Supplementary Material contain the additional Refs.~\cite{ITP_dtLS_Lehtovaara2007, ZapataPRA1998}.

\bibitem{ITP_dtLS_Lehtovaara2007}
L. Lehtovaara, J. Toivanen, and J. Eloranta, 
Solution of time-independent Schr\"odinger equation by the imaginary time propagation method. 
{\it Journal of Computational Physics} {\bf 221}, 148 (2007).

\bibitem{ZapataPRA1998}
I. Zapata, F. Sols, and A. J. Leggett, Josephson effect between trapped Bose-Einstein condensates. 
{\it Phys. Rev. A} {\bf 57}, R28 (1998).

\bibitem{Fischer2006}
R. Schützhold, M. Uhlmann, Y. Xu, U. R. Fischer,
Mean-field expansion in Bose-Einstein condensates with finite-range interactions.
\textit{Int. J. Mod. Phys. B} \textbf{20}, 3555 (2006). 

\bibitem{LimaPelster2011}
A. R. P. Lima and A. Pelster, 
Quantum fluctuations in dipolar Bose gases. 
{\it Phys. Rev. A} {\bf 84}, 041604 (2011).

\bibitem{LimaPelster2012}
A. R. P. Lima and A. Pelster, 
Beyond mean-field low-lying excitations of dipolar Bose gases. 
{\it Phys. Rev. A} {\bf 86}, 063609 (2012). 

\bibitem{Petrov2015_eGPE}
D. S. Petrov, 
Quantum Mechanical Stabilization of a Collapsing Bose-Bose Mixture. 
{\it Phys. Rev. Lett.} {\bf 115}, 155302 (2015). 

\bibitem{WchtlerPRA2016}
F. W\"achtler and L. Santos, 
Quantum filaments in dipolar Bose-Einstein condensates. 
{\it Phys. Rev. A} {\bf 93} 061603 (2016).

\bibitem{Roccuzzo2019}
S. M. Roccuzzo and F. Ancilotto, 
Supersolid behavior of a dipolar Bose-Einstein condensate confined in a tube. 
{\it Phys. Rev. A} {\bf 99}, 041601 (2019). 

\bibitem{TengstrandPRA2021}
M. N. Tengstrand, D. Boholm, R. Sachdeva, J. Bengtsson, and S. M. Reimann,
Persistent currents in toroidal dipolar supersolids. 
{\it Phys. Rev. A} {\bf 103}, 013313 (2021).

\bibitem{RoccuzzoPRL2020}
S. M. Roccuzzo, A. Gallemí, A. Recati, and S. Stringari, 
Rotating a Supersolid Dipolar Gas. 
{\it Phys. Rev. Lett.} {\bf 124}, 045702 (2020).

\bibitem{TanziNature2019}
L. Tanzi, et al., 
Supersolid symmetry breaking from compressional oscillations in a dipolar quantum gas.
{\it Nature} {\bf 574}, 382 (2019).

\bibitem{Tanzi2021}
L. Tanzi, et al. 
Evidence of superfluidity in a dipolar supersolid from nonclassical rotational inertia. 
{\it Science} {\bf 371}, 1162 (2021).

\bibitem{CabreraScience2018}
C. R. Cabrera, L. Tanzi, J. Sanz, B. Naylor, P. Thomas, P. Cheiney, and L. Tarruell, 
Quantum liquid droplets in a mixture of Bose-Einstein condensates. 
{\it Science} {\bf 359} 301 (2018).

\bibitem{SemeghiniPRL2018}
G. Semeghini, G. Ferioli, L. Masi, C. Mazzinghi, L. Wolswijk, F. Minardi, M. Modugno,
G. Modugno, M. Inguscio, and M. Fattori,
Self-Bound Quantum Droplets of Atomic Mixtures in Free Space. 
{\it Phys. Rev. Lett.} {\bf 120} 235301 (2018).

\bibitem{Alaa2022}
A. Ala\~na, N. Antolini, G. Biagioni, I. L. Egusquiza, and M. Modugno,
Crossing the superfluid-supersolid transition
of an elongated dipolar condensate. 
{\it Phys. Rev. A} {\bf 106} 043313 (2022).

\bibitem{BiagioniPRX2022}
G. Biagioni, N. Antolini, A. Ala{\~n}a, M. Modugno, A. Fioretti, C. Gabbanini, L. Tanzi, and G. Modugno,
Dimensional crossover in the superfluid-supersolid quantum phase transition. 
{\it Phys. Rev. X} {\bf 12}, 021019 (2022).

\bibitem{nota1}
Exciting the dynamics with a single Gaussian potential under one of the two central clusters instead of the sinusoidal potential under all clusters, induces the coupling to many dynamical modes, preventing the observation of clear Josephson oscillations. 

\bibitem{note1}
Population and phase of $j$-th cluster are defined as $N_j(t) = \int_{cluster j} dx \ n(x,t)$ and $\theta_j = \int_{cluster j} dx \ \theta(x,t)$, where the integrated 1d densities and phases are $n(x,t) = \int dy\ dz\ |\psi(\textbf{r},t)|^2$ and $\theta(x,t) = \frac{\int dy\ dz\ \phi(\textbf{r},t)\ |\psi (\textbf{r},t)|^2}{\int dy\ dz\ |\psi (\textbf{r},t)|^2}$. The integrals on the transverse directions $y-z$ are performed over the whole box, while on the $x$-direction is performed on the cluster $j$th, specifically on the interval between the two minima in the density modulation outlining the limits of the cluster. We have weighted the 1d phase profile with the density to suppress noise at the boundaries, where the density goes to zero.

\bibitem{PfauNatureGoldstone}
M. Guo, et al., 
The low-energy Goldstone mode in a trapped dipolar supersolid. \textit{Nature} \textbf{574}, 386 (2019).

\bibitem{nota2}
In principle, the interaction and coupling parameters might be directly obtained by replacing the order parameter $\Psi(\vect{r},t) = \sum_{j=1}^{4} \phi(\vect{r}) \psi_j(t)$ in the eGPE, neglecting the LHY correction as well as the time dependence of the cluster spatial wavefunction $\phi(\vect{r})$.
However, employing this approach makes it challenging to accurately discern the cluster wavefunction $\phi(\mathbf{r})$ and obtain trustworthy outcomes.

\bibitem{GatiPRL2006}
R. Gati, B. Hemmerling, J. Fölling, M. Albiez, and M. K. Oberthaler,
Noise Thermometry with Two Weakly Coupled Bose-Einstein Condensates,
{\it Phys. Rev. Lett.} {\bf 96}, 130404 (2006).

\bibitem{MilburnPRA1997}
G. J. Milburn, J. Corney, E. M. Wright and D. F. Walls,
Quantum dynamics of an atomic Bose-Einstein condensate in a double-well potential,
{\it Phys. Rev. A} {\bf 55}, 4318 (1997).

\bibitem{EsteveNATURE2008}
J. Esteve, C. Gross, A. Weller, S. Giovanazzi, and M. K. Oberthaler,
Squeezing and entanglement in a Bose–Einstein condensate
{\it Nature} {\bf 455}, 1216 (2008).

\bibitem{PezzePRA2006}
L. Pezzè, L. A. Collins, A. Smerzi, G. P. Berman, and A. R. Bishop,
Sub-shot-noise phase sensitivity with a Bose-Einstein condensate Mach-Zehnder interferometer,
{\it Phys. Rev. A} {\bf 72}, 043612 (2006).

\bibitem{BerradaNATCOMM2013}
T. Berrada, S. Van Frank, R. Bücker, T. Schumm, J. F. Schaff, and J. Schmiedmayer,
Integrated Mach–Zehnder interferometer for Bose–Einstein condensates,
{\it Nature Communications} {\bf 4}, 2077 (2013).

\bibitem{PezzeRMP2018}
L. Pezzè, A. Smerzi, M. K. Oberthaler, R. Schmied, and P. Treutlein,
Quantum metrology with nonclassical states of atomic ensembles,
{\it Rev. Mod. Phys.} {\bf 90}, 035005 (2018).

\bibitem{AlanaARXIV}
A. Alaña, M. Modugno, P. Capuzzi, and D. M. Jezek,
Self-sustained Josephson dynamics and self-trapping in supersolids,
arXiv.2501.08739.

\end{thebibliography}
\end{document}